\documentclass[12pt]{iopart}
\usepackage{graphicx}
\usepackage{bm}
\begin{document}

\title[Adiabatic piston in a temperature gradient ]{Adiabatic piston in a temperature gradient }

\author{J Javier Brey and Nagi Khalil}

\address{F\'{\i}sica Te\'{o}rica, Universidad de Sevilla, Apartado de Correos 1065,
E-41080 Sevilla, Spain}

\eads{\mailto{brey@us.es} and \mailto{nagi@us.es}}

\begin{abstract}
The steady states of two gases of hard spheres or disks separated by
an adiabatic piston in presence of a temperature gradient are
discussed. The temperature field is generated by two thermal walls at
different temperatures, each of them in contact with one of the
gases. The presence of the piston strongly affects the hydrodynamic
fields, inducing a jump in its vicinity. A simple kinetic theory model
is formulated. Its  predictions are
shown to be in good agreement with molecular dynamics simulation
results. The applicability of the minimum entropy production principle
is analyzed, and it is found that it only provides an accurate
description of the system in the limit of a small temperature
gradient.
\end{abstract}

\noindent{\it Keywords\/}: Non-equilibrium and irreversible
thermodynamics, Kinetic theory, Fluctuation phenomena, random
processes, noise, and Brownian motion

\pacs{05.70.Ln, 05.20.Dd, 05.40.-a.}

\maketitle

The adiabatic piston consists of a container filled with a gas that is
divided into two compartments by a freely moving adiabatic piston
\cite{Ca63}. Usually, the system is isolated and initially prepared
with the two gases in both compartments at different independent
states. Then, the relaxation to equilibrium is studied. It is observed
that the system first converges towards a state of mechanical
equilibrium with both gases having the same pressure. Then, the
pressure fluctuations that are asymmetrical because the temperatures
of the gases are different, drive the system very slowly to thermal
equilibrium \cite{Fe65,GyL06}. Of course, the situation is much more
complex and less understood when the system can not relax to
equilibrium due to some imposed external conditions. Very recently,
several configurations consisting of granular gases separated by an
adiabatic piston have been considered
\cite{BRyvB05,ByK10,ByK11}. Granular gases are inherent
non-equilibrium systems, since no equilibrium state is possible for
them and, for this reason, they have been intensively used to
investigate many fundamental issues in the context of non-equilibrium
statistical mechanics \cite{Du00,Go03}. Nevertheless, it must be
realized that the irreversibility in granular gases has an internal
origin, and it is not related, in principle, to some imposed boundary
or initial conditions, as it is the case in molecular systems.
Actually, as a consequence of the above, gradients and inelasticity
are related in steady granular gases and this leads to a quite
peculiar rheological behavior.

The aim of this paper is to study the steady state of two molecular gases separated by an adiabatic piston, when each of the gases is in contact with a thermal wall at different temperature. Attention will be restricted to the final state of the system, whithout studying the way in which it is reached.  The questions addressed are related with the modification of the hydrodynamic steady profiles because of the presence of the piston and the steady position of the latter.

\begin{figure}
  \centering
  \includegraphics[scale=0.4]{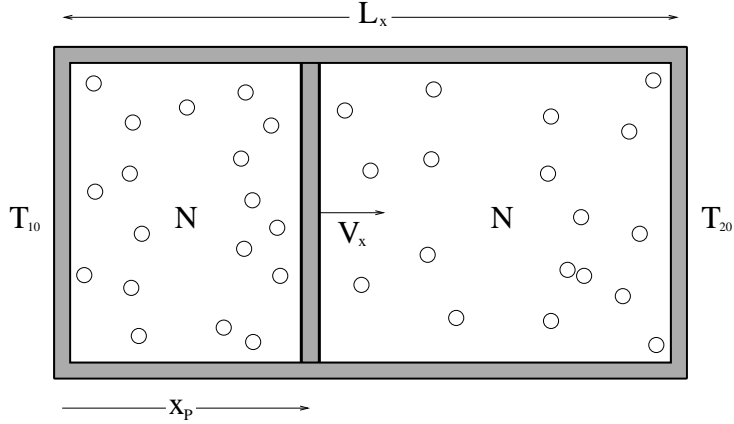}
  \caption{Sketch of the system considered in this work. The left and right walls are thermal with temperatures $T_{10}$ and $T_{20}$, respectively.}
  \label{fig1}
\end{figure}

The system considered is a rectangular ($d=2$) or cylindrical ($d=3$) container of length $L_{x}$. It is divided into two compartments by a movable adiabatic piston of mass $M$, constrained to remain perpendicular to the axis of the system, taken as $x$-axis. By adiabatic it is meant, as usual, that the piston has no internal degrees of freedom and, therefore, it can not transmit energy when it is at rest. A sketch of the system is given in Fig. \ref{fig1}. In each of the two compartment there are $N$ elastic hard disks ($d=2$) or spheres ($d=3$) of mass $m$ and diameter $\sigma$. Collisions between particles and the piston as well as collisions of the particles with the lateral walls of the container are also elastic. Moreover, the motion of the piston occurs without friction with the lateral walls.

The walls of the container located at $x=0$ and $x=L_{x}$ and perpendicular to the $x$-axis are thermal walls with temperatures $T_{10}$ and $T_{20}$, respectively. At the microscopic level of description, a thermal wall is modeled by assuming that whenever a particle collides with it, a new velocity is assigned to the particle drawn from a Maxwellian distribution with a second moment defined by the temperature of the wall \cite{Ce69,DyvB97}. More specifically, the velocity distribution of the  particles leaving a thermal wall with temperature $T_{i0}$, $i=1,2$,
is given by
\begin{equation}
\label{00}
P({\bm v}) =  \theta \left[ (-1)^{i-1} v_{x} \right] \left( 2 \pi \right)^{- \frac{d-1}{2}} \left( \frac{m}{k_{B} T_{i0}} \right)^{\frac{d+1}{2}} v_{x} \exp \left( - \frac{m v^{2}}{2k_{B} T_{i0}} \right),
\end{equation}
where $\theta (x)$ is the Heaviside step function and $k_{B}$ is the Boltzmann constant.
 At the macroscopic level, it will be assumed that the fluid in the vicinity of a thermal wall has the same temperature as the wall.

In the following, attention will be restricted to steady states in which the gases at both sides of the piston are very dilute, so that they verify the local equation of state $p_{i}=n_{i}k_{B}T_{i}$, where $p_{i}$, $n_{i}$, and $T_{i}$, are the local pressure, number density, and temperature of the gas in compartment $ i$, $i=1,2$. Also, in the steady states considered only gradients of the hydrodynamic fields in the $x$ direction are present.

The hydrodynamic Navier-Stokes equation for the velocity field, when applied to the above steady states, implies that the pressure must be uniform in each container. In addition, mechanical equilibrium of the piston requires that the pressure on both faces  be the same,
\begin{equation}
\label{1.1}
p_{1}=p_{2} =p.
\end{equation}
The energy balance in each container reads
\begin{equation}
\label{1.2}
\frac{\partial}{\partial x}\ q_{xi} =0,
\end{equation}
where $q_{xi}$ is the heat flux along the $x$ direction in compartment $i$. According to Fourier law,
\begin{equation}
\label{1.3}
q_{xi}=- \kappa (T_{i}) \frac{\partial T_{i}}{\partial x}\, .
\end{equation}
The thermal conductivity $\kappa(T)$ in the dilute limit being considered is
\begin{equation}
\label{1.4}
\kappa (T)= \frac{d(d+2)^{2}}{16(d-1)}\, \Gamma(d/2) \pi^{-\frac{d-1}{2}} k_{B} \left( \frac{k_{B}T}{m} \right)^{1/2} \sigma^{-(d-1)}\, .
\end{equation}
From Eqs.\ (\ref{1.2})-(\ref{1.4}) it follows that
\begin{equation}
\label{1.5}
\frac{\partial^{2} T_{i}^{3/2}}{\partial x^{2}} =0,
\end{equation}
so that
\begin{equation}
\label{1.6}
T_{i}^{3/2}(x) = a_{i}x+b_{i},
\end{equation}
with $a_{i}$, $b_{i}$ constants. But stationarity also requires that the heat flux be the same in both compartments, since otherwise the energy of the piston could not be constant. Then, $q_{x1}=q_{x2}$ and using the Fourier law this leads to
\begin{equation}
\label{1.7}
a_{1}=a_{2}=a.
\end{equation}
Now the boundary conditions at the thermal walls, $T_{1}(0)=T_{10}$ and $T_{2}(L_{x})=T_{20}$,  are imposed, to get
\begin{equation}
\label{1.8}
T_{1}^{3/2} (x) =a x +T_{10}^{3/2},
\end{equation}
\begin{equation}
\label{1.9}
T_{2}^{3/2}(x)=-a(L_{x}-x)+T_{20}^{3/2}.
\end{equation}
Equation (\ref{1.8}) holds for $0<x<x_{P}$, and Eq.\ (\ref{1.9}) for $x_{P}<x<L_{x}$, where $x_{P}$ is the steady position of the piston.

The boundary conditions associated to the piston must also be introduced. Then, expressions for the energy flux between the piston and the two gases are needed. This flux appears as a consequence of the velocity fluctuations of the piston \cite{ByR09}. Assuming that they are Gaussian in the steady state, something that is confirmed with very good accuracy by the numerical simulations, in the limit $m/M \ll 1$ it is obtained that the energy flux from the piston to the gas in compartment $i$ is given by \cite{ByK10}
\begin{equation}
\label{1.10}
Q_{i} \approx - 2 \left( \frac{2m k_{B}}{\pi} \right)^{1/2} \frac{T_{i}(x_{P})-T_{P}}{T_{i}^{1/2}(x_{P})} \,\frac{p}{M},
\end{equation}
where $T_{P}$ is the temperature parameter of the piston, defined from the second moment of its velocity distribution, and the temperature $T_{i}(x_{P})$ is to be understood as the temperature of the gas in compartment $i$ in the limit $x \rightarrow x_{P}$ taken from inside the gas. Stationarity of the piston yields $Q_{1}+Q_{2}=0$, i.e.
\begin{equation}
\label{1.11}
\frac{T_{1}(x_{P})-T_{P}}{T_{1}^{1/2} (x_{P})} = \frac{T_{2}(x_{P})-T_{P}}{T_{2}^{1/2} (x_{P})}\,
\end{equation}
or
\begin{equation}
\label{1.12}
 T_{P} = \left[ T_{1}(x_{P}) T_{2}(x_{P}) \right]^{1/2}\, .
\end{equation}
A second condition at the piston follows from the continuity of the energy flux, implying that
\begin{equation}
\label{1.13}
q_{x1}(x_{P})=-Q_{1}.
\end{equation}
Use of Eqs. (\ref{1.3}) and (\ref{1.10}) yields
\begin{equation}
\label{1.14}
T_{2}^{1/2}(x_{P})-T_{1}^{1/2}(x_{P}) = C \frac{a}{p},
\end{equation}
where
\begin{equation}
\label{1.15}
C \equiv \frac{d(d+2)^{2} \Gamma \left( d/2 \right)\pi^{-\frac{d-2}{2}}k_{B} M}{48 \sqrt{2} (d-1)m \sigma^{d-1}}\, .
\end{equation}
The last needed condition is that the number of particles in each compartment is fixed to $N$, so it must be
\begin{equation}
\label{1.16}
S_{P} \int_{0}^{x_{P}} dx\, n_{1}(x) = S_{P} \int_{x_{P}}^{L_{x}} dx\, n_{2} (x) = N,
\end{equation}
where $S_{P}$ is the area ($d=3$) or length ($d=2$) of the piston and, therefore, also the section of the container. By employing the local equation of state and Eqs.\ (\ref{1.8}) and (\ref{1.9}), the above conditions (\ref{1.14}) and (\ref{1.16}) are seen to be equivalent to
\begin{equation}
\label{1.17}
T_{1}^{1/2} (x_{P})+T_{2}^{1/2} (x_{P}) = T_{10}^{1/2}+T_{20}^{1/2}
\end{equation}
and
\begin{equation}
\label{1.18}
T_{1}^{1/2}(x_{P}) =T_{10}^{1/2} + D \frac{a}{p}\, ,
\end{equation}
with
\begin{equation}
\label{1.19}
D \equiv \frac{Nk_{B}}{3S_{P}}.
\end{equation}
Equations (\ref{1.14}), (\ref{1.17}), and (\ref{1.18}) form a closed set of equations for the unknown $T_{1}(x_{P})$, $T_{2}(x_{P})$, and $a/p$. Solving it gives
\begin{equation}
\label{1.20}
T_{1}^{1/2}(x_{P})= T_{10}^{1/2} + D\, \frac{T_{20}^{1/2}-T_{10}^{1/2}}{2D+C}\, ,
\end{equation}
\begin{equation}
\label{1.21}
 T_{2}^{1/2}(x_{P})= T_{20}^{1/2} - D\, \frac{T_{20}^{1/2}-T_{10}^{1/2}}{2D+C}\, ,
\end{equation}
\begin{equation}
\label{1.22}
\frac{a}{p}= \frac{T_{20}^{1/2}-T_{10}^{1/2}}{2 D+C}\, .
\end{equation}
The only remaining task is to identify $a$ (or $p$) and the steady position of the piston $x_{P}$. By means of Eqs.\ (\ref{1.8}) and (\ref{1.9}) it is found
\begin{equation}
\label{1.23}
a= \frac{T_{1}^{3/2}(x_{P})-T_{2}^{3/2}(x_{P})-T_{10}^{3/2}+T_{20}^{3/2}}{L_{x}},
\end{equation}
\begin{equation}
\label{1.24}
x_{P}= \frac{L_{x} \left[ T_{1}^{3/2}(x_{P})-T_{10}^{3/2} \right]}{T_{1}^{3/2}(x_{P})-T_{2}^{3/2}(x_{P})-T_{10}^{3/2}+T_{20}^{3/2}}\, .
\end{equation}
Therefore, as a consequence of the presence of the piston the slope $a$ of the temperature profile decreases as compared with its values in a gas without piston, being the same the temperatures of the two thermal walls.

To check the accuracy of the above theoretical predictions, molecular dynamics (MD) simulations of a system of hard disks ($d=2$)  have been performed, using an event driven algorithm \cite{AyT87}. The number of particles in each of the two compartments has been $N=100$ and the size
 of the system  is $L_{x}=2S_{P}= 100 \sigma$. The values of $M/m$ and $T_{20}/T_{10}$ have been varied as indicated below. All the simulations started with the piston located in the middle of the system ($x_{P}=L_{x}/2$) and the particles homogeneously distributed in each compartment. After a short transient time, of the order of a few collisions per particle, a steady state was always reached. Then, density and temperature profiles of the gases in both compartments as well as the position and velocity distribution of the piston were measured. To identify the hydrodynamic profiles, the system was divided in $20$ layers of  the same width, parallel to the piston. Moreover, the average quantities have been averaged on time (typically $200$ registers) and also over a given number of trajectories (typically $50$).

\begin{figure}
\centering
\includegraphics[scale=0.5,angle=-90]{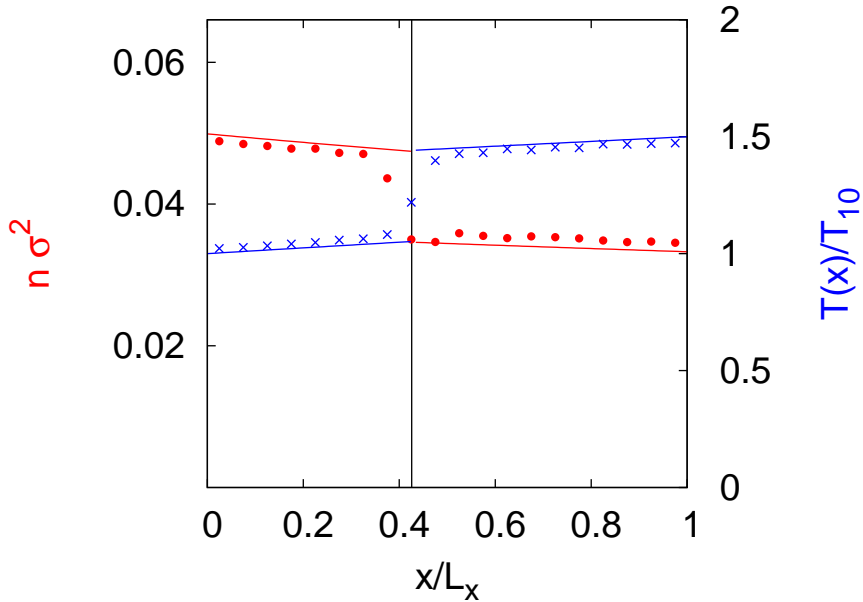}
\caption{Steady temperature (blue crosses and solid line) and density (red circles and solid line) profiles. The symbols are MD simulation results for a tow-dimensional system of $2N=200$ hard disks with $L_{x}=2S_{P}=100 \sigma$, $M/m=10$, and $T_{20}/T_{10}=1.5$. The solid lines are the theoretical predictions derived in the main text.   }
\label{fig2}
\end{figure}

\begin{figure}
\centering
\includegraphics[scale=0.5,angle=-90]{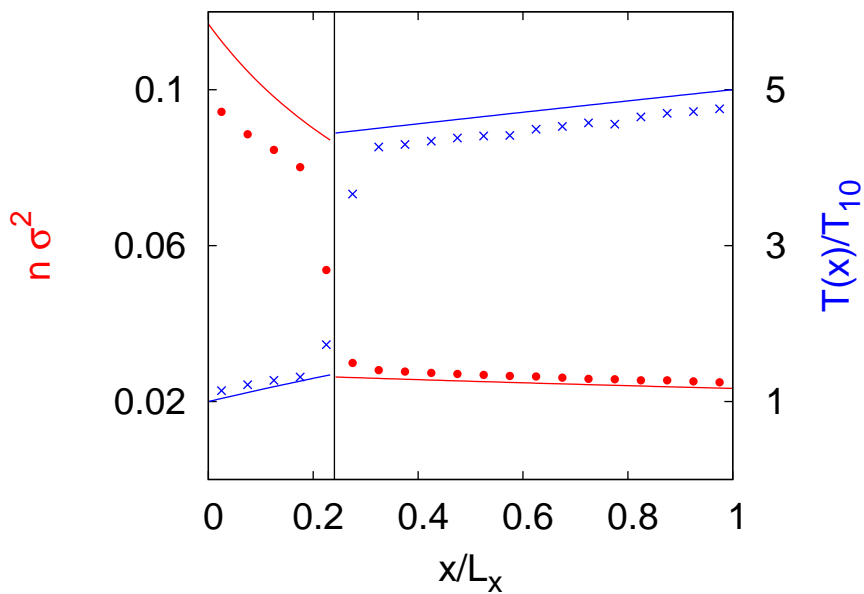}
\caption{The same as in Fig.\ \protect{\ref{fig2}}, but now $T_{20}/T_{10}=5$}
\label{fig3}
\end{figure}

 In Fig.\ \ref{fig2}, the steady temperature and density profiles are plotted for a system with $M/m=10$ and $T_{20}/T_{10}=1.5$. The profiles for a system with the same value of the mass ratio $M/m$, but with a much larger temperature ratio, $T_{20}/T_{10}=5$, are given in Fig. \ref{fig3}. The symbols are MD results, while the solid lines are the theoretical predictions obtained above. The vertical line indicates the average position of the piston. A fairly good agreement is observed. It is important to realize that the hydrodynamic fields in the vicinity of the average position of the piston are much influenced by the fluctuations of the latter, that are not accounted for in the simple model developed here.

\begin{figure}
\centering
\includegraphics[scale=0.5,angle=-90]{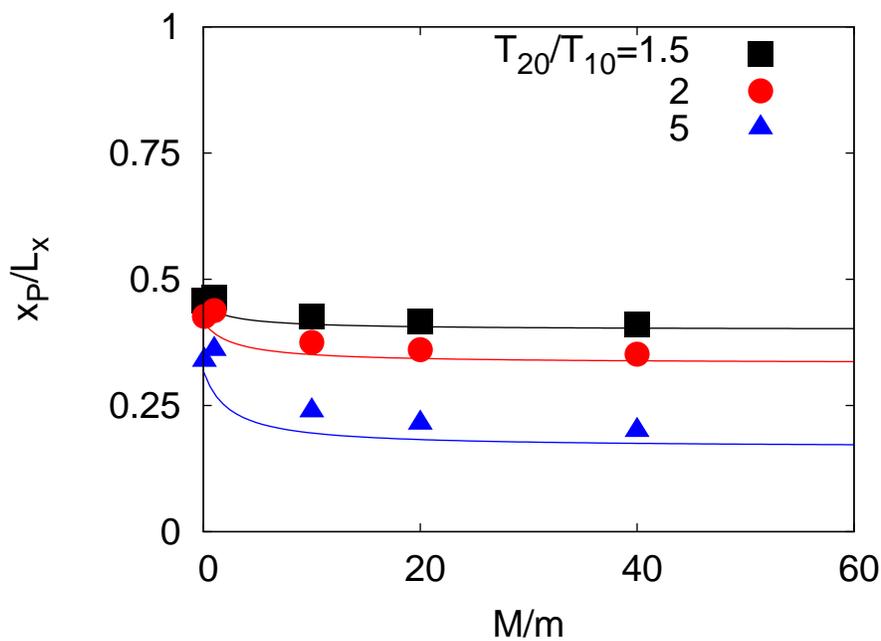}
\caption{Steady position of the piston $x_{P}$ as a function of the mass of the piston $M$. Three values of the ratio of the temperatures of the two thermal walls have been considered, as indicated. The values of $L_{x}$ and $N$ are the same as in Fig. \ref{fig2}. The symbols are MD simulation results while the solid lines are the predictions derived in the main text. }
\label{fig4}
\end{figure}

\begin{figure}
\centering
\includegraphics[scale=0.5, angle=-90]{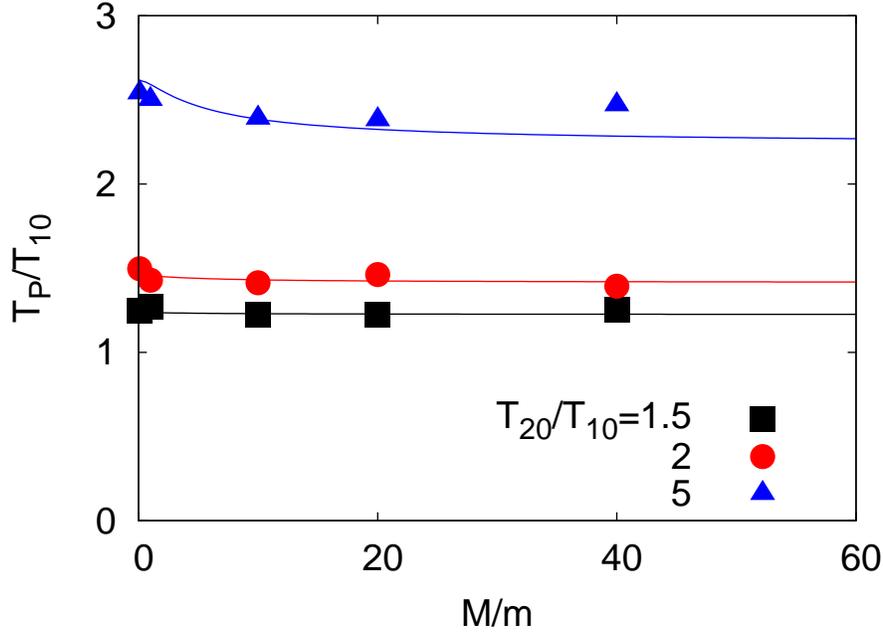}
\caption{The same as in Fig\ \protect{\ref{fig4}}, but now the plotted quantity is the temperature of the piston divided by the temperature of the cold wall.}
\label{fig5}
\end{figure}

The results for the dependence of the steady position of the piston $x_{P}$ on the mass ratio $M/m$ are shown in Fig\ \ref{fig4}. Three different values of the ratio of the temperatures of both thermal wall have been considered, as indicated in the inset. As the mass ratio increases the position of the piston tends to a constant value. A similar behavior is observed for the temperature of the piston, shown in Fig. \ref{fig5}. In both cases, there is again a fairly good agreement between the simulation results and the theoretical prediction from the model proposed here.

It seems worth to investigate whether the steady state reached by the system considered here can be described by means of a minimum entropy production rule. The entropy production function $P$ of the system is \cite{dGyM62,McL89}
\begin{eqnarray}
\label{1.25}
P & = & S_{P} \int_{0}^{x_{P}} dx\, q_{x1}(x) \frac{dT^{-1}_{1}(x)}{dx}
\nonumber \\
&& + S_{P} \int_{x_{P}}^{L_{x}} dx\, q_{x1}(x) \frac{dT_{2}^{-1}(x)}{dx} \nonumber \\
&& - Q_{1} \left[ \frac{1}{T_{P}}- \frac{1}{T_{1}(x_{P})} \right] \nonumber \\
& & + Q_{2} \left[ \frac{1}{T_{2}(x_{P})} - \frac{1}{T_{P}} \right],
\end{eqnarray}
where the possible discontinuity of the temperature at the piston has been taken into account. In this expression, the relations given in Eqs. (\ref{1.1}), (\ref{1.8}), (\ref{1.9}), and ({\ref{1.16}) are assumed to be known since they correspond to boundary conditions. Moreover, the limit of small temperature gradients will be considered in the following. Define
\begin{equation}
\label{1.26}
T_{m} \equiv \frac{T_{10}+T_{20}}{2}\, , \quad \quad \theta \equiv \frac{T_{20}-T_{10}}{2T_{m}},
\end{equation}
and introduce the scaled temperature deviations $\tau_{1}$, $\tau_{2}$, and $\tau_{P}$ by
\begin{equation}
\label{1.27}
 T_{1}(x_{P}) = T_{m} + \theta \tau_{1}, \quad  T_{2}(x_{P}) = T_{m} + \theta \tau_{2},
 \end{equation}
\begin{equation}
\label{1.27a}
 \quad  T_{P} = T_{m} + \theta \tau_{P}.
\end{equation}
When these expressions are substituted into Eq.\ (\ref{1.25}) and  an expansion in powers of $\theta $ is carried out, it is obtained that
\begin{eqnarray}
\label{1.28}
P(\tau_{1}, \tau_{2}, \tau_{3}) & = & \frac{6 (2mk_{B})^{1/2} S_{P}}{\pi^{1/2} M L_{x} T_{m}^{3/2}} \nonumber \\
&&\times \left\{ 2D \left[\tau_{1}^{2} + \tau_{2}^{2}-2 \left( \tau_{1}+ \tau_{2} \right) \tau_{P} + 2 \tau_{P}^{2} \right] \right. \nonumber \\
&& + \left. C \left[ \tau_{1}^{2}+ \tau_{2}^{2} +2 T_{m} (\tau_{1}- \tau_{2} +T_{m}) \right] \right \} \theta^{2} \nonumber \\
 && + \mathcal{O} ( \theta^{4}).
\end{eqnarray}
The values of $\tau_{1}$, $\tau_{2}$ and $\tau_{3}$ minimizing this expression for small $\theta$, namely neglecting terms of order $\theta^{4}$ and higher, are
\begin{equation}
\label{1.29}
\tau_{1} =- \frac{CT_{m}}{C+2D}\, , \quad  \tau_{2} = - \tau_{1}, \quad
\tau_{P}=0.
\end{equation}
Using these values, it is easily seen that
\begin{equation}
\label{1.30}
x_{P} = \frac{L_{x}}{2} \left( 1-\frac{C+D}{C+2D}\, \theta \right) , \quad p= \frac{2Nk_{B}T_{m}}{L_{x}S_{P}}\, ,
\end{equation}
\begin{equation}
\label{1.30a}
\quad a_{1}=a_{2}= \frac{6D}{C+2D}\frac{T_{m}^{3/2} \theta}{L_{x}}.
\end{equation}
These results agree with the lowest order expansion in $\theta$ of Eqs.\ (\ref{1.22})-(\ref{1.24}), as it can be easily verified. On the other hand, it is also easily seen that the condition that the entropy production $P$ is minimum is not equivalent to the hydrodynamic theory developed here  outside of the limit of small temperature gradients, leading to results that strongly disagree with the numerical simulations.

In summary, it has been shown that the presence of the adiabatic piston introduces a strong discontinuity of the hydrodynamic profiles, increasing their slope in both compartments. Moreover. the simple model presented here based on the energy balance, reproduces quite well the molecular dynamics simulation results. A minimum entropy production requirement only holds in the limit of small temperature differences. Although we are not aware of any  realization of a situation similar to the one considered here, the effects discussed  should be observable in experiments, by extending the devices used to investigated the relaxation to equilibrium \cite{Ru29,PyL97,LyP98}.

\ack
This research was supported by the Ministerio de Educaci\'{o}n y Ciencia (Spain)
through Grant No. FIS2011-24460 (partially financed by FEDER funds).


\section*{References}
\begin{thebibliography}{0}

\bibitem{Ca63} Callen H B  1963 {\it Thermodynamics} (New York: Wiley
  and Sons)

\bibitem{Fe65} Feynman R P 1965 {\it Lectures Notes in Physics} (New
  York: Addison-Wesley)

\bibitem{GyL06} Gruber Ch and Lesne A  2006 {\it Encyclopedia of
    Mathematical Physics} ed J P Francoise, G Naber and T S Tsun
  (Amsterdam: Elsevier) p~160

\bibitem{BRyvB05} Brito R, Renne M J and Van der Broeck C 2005
  Europhys. Lett. {\bf 70} 29

\bibitem{ByK10} Brey J J and Khalil N 2010 Phys. Rev. E {\bf 82}
  051301

\bibitem{ByK11} Brey J J  and Khalil N  2011 Europhys. Lett {\bf 94}
  14003

\bibitem{Du00} Dufty J W 2000 J. Phys.: Condens. Matter {\bf 12} A47

\bibitem{Go03} Goldhirsch I (2003) Annu. Rev. Fluid Mech. {\bf 35} 267

\bibitem{Ce69} Cercignani C 1969 Mathematical Methods in Kinetic
  Theory (New York: Plenum)

\bibitem{DyvB97} Dorfman J R and van Beijeren H 1997 Statistical
  Mechanics Part B ed  B J Berne (New York: Plenum) p~65

\bibitem{ByR09} Brey J J and Ruiz-Montero M J 2009 Phys. Rev. E {\bf
    79} 031305

\bibitem{AyT87} Allen M P and Tisdesley D J 1987 Computer Simulation
  of Liquids (New York: Oxford Science Publications)

\bibitem{dGyM62} De Groot S R and Mazur P  1962 Non-Equilibrium
  Thermodynamics (Amsterdam: North-Holland)

\bibitem{McL89} McLennan J A 1989 Introduction to Non-Equilibrium
  Statistical Mechanics (London: Prentice Hall)

\bibitem{Ru29} R\"{u}chardt E 1929 Z. Phys. {\bf 30} 58

\bibitem{PyL97} Pierrus J and de Lange O L  1997 Phys. Rev. E {\bf 56}
  2841

\bibitem{LyP98} de Langue O L and Pierrus J 1998 Phys. Rev. E {\bf 57}
  5520

\end{thebibliography}
\end{document}